\documentclass{jps-cp}
\usepackage{txfonts} 

\title{Deuteron analyzing powers $A_y$, $A_{yy}$ and $A_{xx}$ in $dp$-
elastic scattering at large transverse momenta}

\author{V.P.~\textsc{Ladygin}$^{1}$, A.V.~\textsc{Averyanov}$^1$,  E.V.~\textsc{Chernykh}$^1$, 
D.~\textsc{Enache}$^2$, Yu.V.~\textsc{Gurchin}$^1$, A.Yu.~\textsc{Isupov}$^1$, M.~\textsc{Janek}$^3$, 
J.-T.~\textsc{Karachuk}$^{1,2}$, 
A.N.~\textsc{Khrenov}$^1$, D.O.~\textsc{Krivenkov}$^1$, P.K.~\textsc{Kurilkin}$^1$, N.B.~\textsc{Ladygina}$^1$,
A.N.~\textsc{Livanov}$^1$, O.~\textsc{Mezhenska}$^4$, S.M.~\textsc{Piyadin}$^1$,  S.G.~\textsc{Reznikov}$^1$,
Ya.T.~\textsc{Skhomenko}$^{1,5}$, A.A.~\textsc{Terekhin}$^1$, A.V.~\textsc{Tishevsky}$^1$,    T.~\textsc{Uesaka}$^6$ and I.S.~\textsc{Volkov}$^{1,5}$ (DSS Collaboration)}

\inst{$^{1}$Joint Institute for Nuclear Research, Dubna, Russian Federation\\
$^{2}$National Institute for R\&D in Electrical Engineering ICPE-CA, Bukharest, Romania\\
        $^3$  Physics Department, University of \v{Z}ilina, \v{Z}ilina, Slovak Republic\\
        $^4$ P.-\v{J}.~Shafarik University, Koshi\v{c}e, Slovak Republic\\
        $^5$ Belgorod State National Research University, Belgorod, Russian Federation\\
        $^6$ Nishina Center for Accelerator-Based Science, RIKEN,  Wako,   Japan}

\email{vladygin@jinr.ru}

\recdate{December 21, 2021}

\abst{The results on the  the vector $A_y$ and 
tensor $A_{yy}$ and $A_{xx}$ analyzing powers in deuteron-proton 
elastic scattering at large scattering angles are presented. 
These data were  obtained at internal target at JINR Nuclotron 
in the energy range 400-1800 MeV using polarized deuteron beam 
from new polarized ion source.
 New  data on the deuteron analyzing powers 
in the wide energy range  demonstrate the sensitivity to the short-range spin structure 
of the isoscalar nucleon-nucleon correlations.
}

\kword{short-range correlations, spin structure, analyzing powers, polarization} 

\begin{document}
\maketitle

\section{Introduction}

The short-range correlations (SRC) of nucleons in nuclei is the subject of intensive 
theoretical and experimental works during last years.  
Since SRC have densities comparable to
the density in the center of a nucleon which is about $\rho\sim 5\rho_0$ 
($\rho_0\approx 0.17$~fm$^{-3}$), they can be considered as the  
drops of {cold dense nuclear matter} \cite{strikman1}.  
The results obtained at BNL \cite{strikman2}, SLAC \cite{dday} and  JLAB \cite{CLAS_inc, CLAS_2N_3N} 
clearly demonstrate that  more than 90\% all nucleons with momenta
$k\ge 300$~MeV/$c$ belong to 2N SRC; the  probability for a given proton with momenta
$300\le k\le 600$~MeV/$c$ to belong to $pn$ correlation is $\sim$18 times larger than
for $pp$ correlations; the  probability for a nucleon to have momentum $\ge 300$~MeV/$c$
in medium nuclei is $\sim$25\%;  3N SRC are present in nuclei with a significant 
probability \cite{strikman3}. However, still many open questions persist and further
investigations are required both from the experimental and theoretical sides. 

The main goal of the Deuteron Spin Structure (DSS) experimental program is to obtain the information on the  spin - dependent parts of two-nucleon ($2N$) and  three-nucleon ($3N$) forces from two processes: $dp$- elastic scattering in a wide energy range and $dp$- nonmesonic breakup with two protons detection at energies 300 -- 500 MeV 
\cite{dss1,dss2,dss3,lad1}  using the Nuclotron internal target station (ITS) \cite{ITS}. 
Such experimental program at Nuclotron was started by the measurements of the vector 
$A_y$ and tensor $A_{yy}$ and $A_{xx}$ analyzing powers  in $dp$- elastic scattering at $T_d$ of 880~MeV \cite{PLB2012} and 2000~MeV \cite{PPNL2011}. The systematic measurements of the differential cross section  have been performed also in recent years 
\cite{gurchin2013,terekhin2015, terekhin2017, terekhin2019}. 

In this paper we report new results of the  energy scan of the vector $A_y$ and tensor $A_{yy}$ and $A_{xx}$ analyzing powers
in $dp$- elastic scattering obtained at the Nuclotron ITS \cite{ITS} in the energy range of 400-1800~MeV.

\section{Experiment at Nuclotron ITS}

The ITS setup is well suited for study of the energy dependence of
polarization observables for the deuteron-proton elastic scattering and deuteron 
breakup reaction with the detection of two protons at large scattering angles.  
For these purposes the CH$_2$-target of 10 $\mu$m thick is used for the measurements.
The yield from  carbon content of the CH$_2$-target is estimated in separate measurements using several twisted 8$\mu$m carbon wires.  The monitoring of the intensity is done from the detection of 
$pp$- quasielastic scattering at $90^\circ$ in cms by the scintillation counters placed 
in the horizontal plane.
The detection of the $dp$- elastic events is done by the coincidence measurements 
of the proton and deuteron. The detectors are placed in the both horizontal and vertical
planes for the analyzing powers measurements. The selection of the $dp$- elastic events
is done by the correlation of the energy losses in plastic scintillators 
for deuteron and proton and their time-of-flight difference.  
The use of large amount of the scintillation counters allowed to cover
wide angular range \cite{ITS_polarimeter}. 
Such a method has been used to
obtain the polarization data in $dp$- elastic scattering at $T_d$ of 880~MeV \cite{PLB2012} and 2000~MeV \cite{PPNL2011}.

The upgraded setup at ITS  has been used to measure the vector $A_y$ and tensor $A_{yy}$ and $A_{xx}$ 
analyzing powers in $dp$- elastic scattering between 400 MeV and 1800 MeV  
using polarized deuteron beam from 
new source of polarized ions (SPI) developed at LHEP-JINR \cite{NewPIS}.
These measurements were performed using internal target station at Nuclotron \cite{ITS} with new   
control and  data  acquisition system \cite{ITS_DAQ}. 
The existing setup \cite{ITS_polarimeter} has been upgraded 
by new VME based DAQ, \cite{isupov_dspin2017},
new MPod based high voltage system, \cite{skhomenko_mpod}, 
new system of the luminosity monitors etc. 
The same setup has been used as a polarimeter based on the use of  $dp$- elastic scattering at large angles 
($\theta_{\rm cm}\ge 60^{\circ}$) at 270 MeV\cite{ITS_polarimeter}. 

\section{Measurements of the analyzing power in $dp$- elastic scattering}
 
New SPI \cite{NewPIS} has been used to provide 
polarized deuteron beam. In the current experiment the spin modes with the maximal ideal values 
of ($P_z$,$P_{zz}$)= (0,0), (+1/3,+1) and (+1/3, +1) were used. 
The deuteron beam polarization has been measured at 270 MeV \cite{ITS_polarimeter} where
precise values of the deuteron analyzing powers exist \cite{dp_270}.
The $dp$- elastic scattering  events at 270 MeV were selected using correlation of the energy losses 
and time-of-flight difference  for deuteron and proton detectors.  
The values of the beam polarization for different spin  have been obtained  as weighted averages for
8 scattering angles for $dp$- elastic scattering in the horizontal plane only. The typical
values of the beam polarization were $\sim$65-75\% from the ideal values.

After deuteron beam polarization measurements at 270 MeV, the beam has been accelerated up to the required energy
$T_d$ between 400 MeV and 1800 MeV.
The scintillation detectors were positioned in the horizontal and vertical plane 
in accordance with the kinematic of $dp$- elastic scattering for the investigated energy 
The main part of the measurements were performed using 
CH$_2$ target. Carbon target was used  to estimate the background.
The selection of the $dp$- elastic events
is done by the correlation of the energy losses in plastic scintillators 
for deuteron and proton and their time-of-flight (TOF) difference.   
The normalized numbers of $dp$-elastic 
scattering events for each spin mode were used to calculate the values of 
the analyzing powers
$A_y$, $A_{yy}$ and $A_{xx}$.

\begin{figure}[htbp]
\centering
{\includegraphics[width=10cm,clip]{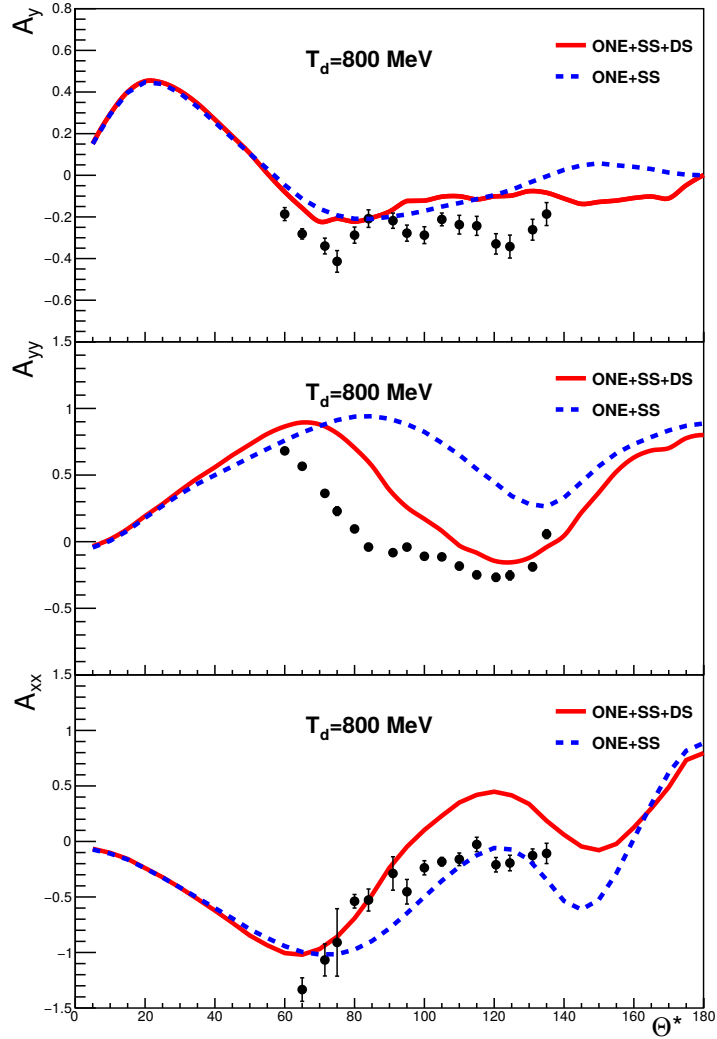}}
\caption{The angular dependencies  of the vector   $A_{y}$ and tensor $A_{yy}$ and $A_{xx}$ analyzing powers at the deuteron kinetic energy $T_d$ of 800~MeV. The full circles are
the results of the experiment performed at ITS at Nuclotron. Predictions of relativistic multiple scattering model with one nucleon
exchange and single scattering (ONE+SS) terms are represented by dashed curves and with additional double scattering (DS)
term by solid ones. The errors are statistical only.}
\label{fig:fig1}
\end{figure}

The preliminary results on the vector $A_y$ and tensor $A_{yy}$ and $A_{xx}$ analyzing powers at the deuteron kinetic energy $T_d$ of 800~MeV are  presented in
Fig.~\ref{fig:fig1}. Predictions of relativistic multiple scattering model \cite{nadia2008,nadia2009} with one nucleon
exchange and single scattering (ONE+SS) terms are represented by dashed curves and with additional double scattering (DS) term by solid ones.
Note that the contribution of the $\Delta$-
isobar mechanism is negligible at these energies \cite{nadia2016,nadia2020}.
The relativistic multiple scattering model \cite{nadia2008, nadia2009} describes the $A_y$ data satisfactorilly,
while it fails to reproduce the tensor analyzing powers data. The considering of the DS term does not improve the agreement.

\begin{figure}[htbp]
\centering
{\includegraphics[width=10cm,clip]{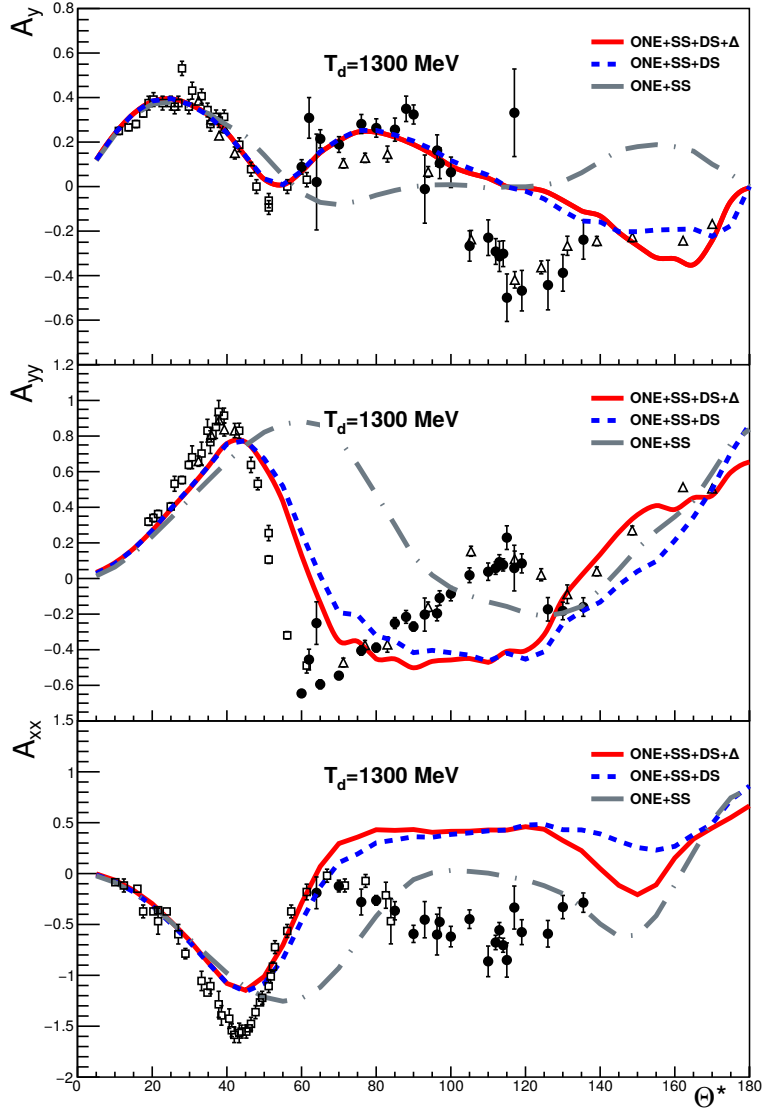}}
\caption{The angular dependencies  of the vector   $A_{y}$ and tensor $A_{yy}$ and $A_{xx}$ analyzing powers at the deuteron kinetic energy $T_d$ of 1300~MeV. The full circles are
the results of the experiment performed at ITS at Nuclotron. 
Open triangles and squares are the data obtained at 1200~MeV at Saclay 
\cite{arv1200} and at ANL \cite{anl_1200}, respectively.  
 Predictions of relativistic multiple scattering model with one nucleon
exchange and single scattering (ONE+SS) terms are represented by dot dashed curves and with additional double scattering (DS)
term by dashed ones. Solid curves represent calculations which include $\Delta$- isobar contribution \cite{nadia2020}. The errors are statistical only.}
\label{fig:fig2}
\end{figure}

The preliminary results on the vector $A_y$ and tensor $A_{yy}$ and $A_{xx}$ analyzing powers at the deuteron kinetic energy $T_d$ of 1300~MeV are  presented in
 Fig.~\ref{fig:fig2}. Open triangles and squares are the data obtained at 1200~MeV at Saclay 
\cite{arv1200} and at ANL \cite{anl_1200}, respectively. One can see a good agreement of the data obtained at Nuclotron with the data from previous experiments \cite{arv1200, anl_1200}. 
The lines are results of
the theoretical calculations  obtained   
in the relativistic multiple scattering expansion formalism \cite{nadia2016,nadia2020}.
The four contributions are taken into account: one-nucleon-exchange (ONE), single- and double-  scattering (SS and DS), and $\Delta$-
isobar excitation. The presented approach was applied earlier to describe the differential cross sections at deuteron
energies between 500 and 1300 MeV in a whole angular range \cite{nadia2016}. 
The dash-dotted, dashed and solid lines are the predictions obtained 
within relativistic multiple scattering model \cite{nadia2016} considering 
ONE+SS terms only, with the DS contribution and with $\Delta$-
isobar excitation term, respectively.
 One can see that the model describes the behavior of 
the vector analyzing power $A_y$ up to $\sim$100$^\circ$ in cms, while the
tensor analyzing powers $A_{yy}$ and $A_{xx}$ are not described over  whole
range of measurements. The $\Delta$- isobar excitation term gives a significant
contribution at the angles larger than 140$^\circ$ in cms.
Apparently, spin structure of the nucleon-nucleon interactions and deuteron at
short distances is missed in the standard description used in the relativistic multiple scattering model \cite{nadia2008,nadia2009,nadia2016,nadia2020}.

The availability of the polarized proton beam at Nuclotron \cite{lad_pp} allows to extend the DSS physics program at ITS \cite{lad1}, namely, to perform  the experiments on the measurements of the nucleon analyzing power $A_y^p$ in  $pd$- elastic scattering   at 135-1000 MeV, in $pd$- nonmesonic breakup at the energies between  135-250 MeV for different kinematic configurations etc.

\section{Conclusions}

The energy scan of the deuteron analyzing powers $A_y$, $A_{yy}$ and $A_{xx}$ in $dp$- elastic scattering has been performed using polarized deuteron beam from new SPI \cite{NewPIS} at upgraded JINR-Nuclotron. 
The data demonstrate the sensitivity to the short-range spin structure of the deuteron.

Next experiments using polarized deuterons and protons at ITS \cite{ITS,ITS_polarimeter} are in  preparation.

The authors thank the Nuclotron staff for providing good conditions of the experiment. They 
thank A.S.~Belov, V.B.~Shutov and V.V~Fimushkin  for the tune of the SPI \cite{NewPIS}. 
They express the gratitude to
S.N.~Bazylev, V.I.~Maximenkova, I.V.~Slepnev, V.M.~Slepnev and  A.V.~Shutov for the help 
during the preparation of the detector and DAQ system.
The work has been supported in part by the RFBR under grant  $N^0$19-02-00079a,  by APVV-20-0052, by
JINR- Slovak Republic and JINR-Romania scientific cooperation programs in 2019-2021.


\begin{thebibliography}{99}
\bibitem{strikman1}  L.~Frankfurt, M.~Sargsian, and M.~Strikman,
{Int.J.Mod.Phys.} \textbf{A23},  2991 (2008).

\bibitem{strikman2} E.~Piasetzky, M.~Sargsian, L.~Frankfurt, M.~Strikman, and J.W.~Watson,
{Phys.Rev.Lett.} \textbf{97}, 162504 (2006).
\bibitem{dday} L.L.~Frankfurt, M.I.~Strikman, D.B.~Day, and M.M.~Sargsian, 
{Phys.Rev. C} \textbf{48}, 2451 (1993).
\bibitem{CLAS_inc} K.Sh.~Egiyan \textit{et al.},
{Phys.Rev. C} \textbf{68},  014313 (2003).
\bibitem{CLAS_2N_3N} K.S.~Egiyan \textit{et al.},
{Phys.Rev.Lett.} 
\textbf{96}, 082501 (2006).
\bibitem{strikman3} L.~Frankfurt, M.~Sargsian, and M.~Strikman,
{AIP Conf.Proc.} \textbf{1056}, 322  (2008). 
\bibitem{dss1} V.P.~Ladygin  \textit{et al.}, 
{Phys.Part.Nucl.} \textbf{45}, 327 (2014).
\bibitem{dss2} V.P.~Ladygin  \textit{et al.}, 
{Few Body Syst.} \textbf{55}, 709 (2014).
\bibitem{dss3} M.~Janek  \textit{et al.}, 
{Few Body Syst.} \textbf{58}, 40 (2017).
\bibitem{lad1} V.P.~Ladygin  \textit{et al.},
{Int. J. Mod. Phys. Conf. Ser.} \textbf{40}, 1660074 (2016).  
\bibitem{ITS} 
A.I.~Malakhov \textit{et al.},   
{Nucl.Instrum.Meth. in Phys.Res.~A}  \textbf{440},  320 (2000).
\bibitem{PLB2012}
P.K.~Kurilkin \textit{ et al.},
{Phys.Lett.~B} \textbf{715},  61 (2012).
\bibitem{PPNL2011}
P.K.~Kurilkin \textit{et al.}, 
{Phys.Part.Nucl.Lett.} \textbf{8},  1081 (2011). 

\bibitem{gurchin2013} Yu.V.~Gurchin \textit{et al.},
{Phys.Part.Nucl.Lett.} \textbf{10}, 243 (2013).
\bibitem{terekhin2015}  A.A.~Terekhin \textit{et al.},
{Phys.Part.Nucl.Lett.} \textbf{12},  695 (2015).  
\bibitem{terekhin2017} A.A.~Terekhin \textit{et al.},
{Phys.Atom.Nucl.} \textbf{80}, 1061 (2017).  
\bibitem{terekhin2019} A.A.~Terekhin \textit{et al.},
{Eur.Phys.J.~A} \textbf{55}, 129 (2019). 
\bibitem{ITS_polarimeter}  P.K.~Kurilkin \textit{et al.},  
{Nucl.Instrum.Meth. in Phys.Res.~A} \textbf{642}, 45 (2011).

\bibitem{NewPIS} A.S.~Belov \textit{et al.}, 
{J.Phys.Conf.Ser. } \textbf{938},  012017  (2017).  


\bibitem{ITS_DAQ}
A.Yu.~Isupov, V.A.~Krasnov, V.P.~Ladygin, S.M.~Piyadin, and S.G.~Reznikov,    
{Nucl.Instrum.Meth. in Phys.Res.~A} \textbf{698}, 127 (2013).
\bibitem{isupov_dspin2017} A.Yu.~Isupov, 
{EPJ Web Conf.} \textbf{204}, 10003 (2019).  
\bibitem{skhomenko_mpod} Ya.T.~Skhomenko \textit{et al.},
{J.Phys.Conf.Ser.} \textbf{938}, 012022 (2017).  
\bibitem{dp_270} K.~Sekiguchi \textit{et al.},
{Phys.Rev.~C} \textbf{65},  034003 (2002);  

K.~Sekiguchi \textit{et al.},
{Phys.Rev.~C} \textbf{70}, 014001 (2004).

\bibitem{nadia2008}  N.B.~Ladygina, {Phys.Atom.Nucl.}  \textbf{71}, 2039 (2008).
\bibitem{nadia2009}  N.B.~Ladygina, {Eur.Phys.J.~A}   \textbf{42}, 91 (2009).
 
\bibitem{nadia2016}  N.B.~Ladygina, 
{Eur.Phys.J.~A} \textbf{52}, 199 (2016). 
\bibitem{nadia2020}  N.B.~Ladygina, 
{Eur.Phys.J.~A} \textbf{56}, 133 (2020). 


\bibitem{arv1200} 
J.~Arvieux \textit{et al.}, Nucl.Phys. A \textbf{431}, 613 (1984).

\bibitem{anl_1200} M.~Haji-Saied  \textit{et al.}, 
{Phys.Rev.~C} \textbf{36},  2010 (1987).


\bibitem{lad_pp} V.P.Ladygin \textit{et al.}, 
{J.Phys.Conf.Ser.} \textbf{938},  012008 (2017).

\end{thebibliography}
\end{document}